\documentclass[journal,10pt,letterpaper]{IEEEtran}
\usepackage{cite}
\usepackage{amsmath,amssymb,mathtools,graphicx,booktabs}
\usepackage{array,tabularx}
\usepackage[hidelinks]{hyperref}
\newcommand{\empavg}[1]{\mathbb E_N\!\left[#1\right]}
\newtheorem{thm}{Theorem}
\newtheorem{lem}[thm]{Lemma}
\newtheorem{prop}[thm]{Proposition}
\newtheorem{cor}[thm]{Corollary}
\newtheorem{rem}[thm]{Remark}
\newtheorem{prob}{Problem}
\newtheorem{defn}{Definition}
\newtheorem{assum}{Assumption}
\newenvironment{pf}{\begin{IEEEproof}}{\end{IEEEproof}}
\begin{document}
\title{Distributed Shaping of Robot Swarms Through Spatial Moments of Arbitrary Order}
\author{Jin Chen and Hector Garcia de Marina%
\thanks{J. Chen and H. Garcia de Marina are with the Department of Computer Engineering, Automation, and Robotics, and the Institute of Mathematics (IMAG) at the University of Granada, Spain. E-mails: \texttt{jinchen920414@gmail.com}; \texttt{hgdemarina@ugr.es}. This work was supported by the ERC Starting Grant \emph{iSwarm} (101076091) and the ERC Proof of Concept Grant \emph{Aerosense} (101289893).}}
\maketitle
\begin{abstract}
This paper shapes a swarm into a target pattern, such as a figure taken from an image, by controlling its spatial moments. The image is sampled into a reference configuration whose centroid and covariance set the position and principal spreads of the formation, while its higher-order moments, expressed in covariance-normalized (whitened) coordinates, encode the remaining shape. In contrast with geometric formation control, which prescribes distances, bearings, or relative positions between specific agents, this specification concerns the swarm as a whole and assigns no agent to a particular place. The key mechanism is orthogonalization in the swarm's own empirical inner product: projecting negative-gradient velocities onto the complement of low-degree polynomials sampled at the agents makes every higher-order correction preserve all lower-order moments exactly. The resulting hierarchical controller has lower triangular error dynamics and converges locally exponentially for any positive gains, so orders can be added without undoing those already achieved. A joint alternative regulates the higher moments together through a regularized Gauss--Newton step while preserving the centroid and covariance dynamics. We give bounds on the number of agents and sufficient conditions for regularity. Both controllers admit a distributed implementation that uses only relative positions and neighbor communication and retains local exponential convergence when estimation is sufficiently fast. In a distributed simulation on a sparse graph, $350$ agents form an airplane from an unstructured swarm through moments up to order ten; the hierarchy reaches the full specification, while the joint law approaches it with smaller velocities.
\end{abstract}
\begin{IEEEkeywords}
Multi-agent systems, swarm robotics, formation control, moment control, distributed estimation, polynomial projection.
\end{IEEEkeywords}
\section{Introduction}\label{sec:introduction}
\IEEEPARstart{F}{ormation} control usually prescribes where each agent should be relative to the others, through relative displacements, distances, bearings, or angles \cite{oh2015survey,fax2004,anderson2008rigid,krick2009,zhao2016bearing,jing2020angle}.
Many collective tasks are instead specified by how the group is distributed in space: a sensing swarm may need to spread more along one direction than another, concentrate toward one side of a region, or take on a recognizable shape, such as a figure taken from an image, regardless of which agent occupies which location.
Prescribing individual destinations then imposes agent-specific constraints that the task does not require, which raises the question of which collective quantities should serve as formation variables and how agent-level feedback can regulate them.

\begin{table*}[t]
\caption{Comparison of distribution-level formation control approaches}
\label{tab:approach-comparison}
\centering
\footnotesize
\setlength{\tabcolsep}{4pt}
\renewcommand{\arraystretch}{1.15}
\begin{tabularx}{\textwidth}{@{}>{\raggedright\arraybackslash}p{0.16\textwidth} *{5}{>{\raggedright\arraybackslash}X}@{}}
\toprule
Approach & Control target & Implementation & Required information & Architecture & Cross-order preservation \\
\midrule
Density feedback \cite{krishnan2018,zheng2022}
& Density profile
& Density-error feedback
& Measured / estimated density; localization varies
& Distributed / mean-field
& Not explicitly enforced \\
\addlinespace
PDE-constrained optimization \cite{sinigaglia2022,sinigaglia2025}
& Density profile
& PDE-constrained optimal control
& Density model; position for field execution\textsuperscript{a}
& Global design; execution varies
& Not explicitly enforced \\
\addlinespace
Second-order moment control \cite{belta2004,freeman2006,yang2008}
& Centroid and covariance (inertial moments)
& Abstraction-based or gradient feedback; consensus estimation
& Absolute positions in a common frame; neighbor messages
& Centralized abstraction / distributed
& No higher orders \\
\addlinespace
Image-moment control \cite{liu2024}
& Higher-order image moments (Legendre, pseudo-Zernike) on a fixed domain
& Gradient of weighted moment error; optimization-based estimation
& Global position in a common frame; neighbor messages
& Distributed execution
& Not enforced; orders weighted empirically \\
\addlinespace
Proposed moment-based control
& Finite-order central moments, covariance-normalized; any dimension
& Projected feedback; distributed estimation
& Relative positions in local frames with common orientation; neighbor messages
& Distributed execution
& All lower orders (hierarchical); centroid and covariance (joint) \\
\bottomrule
\end{tabularx}
\vspace{3pt}
\begin{minipage}{\textwidth}
\scriptsize
\textsuperscript{a}Spatial velocity-field execution in \cite{sinigaglia2025}.
\end{minipage}
\end{table*}

Distribution-level approaches answer this question in different ways (Table~\ref{tab:approach-comparison}).
Density-feedback methods drive the swarm with measured or estimated densities, using local density measurements \cite{krishnan2018}, mean-field feedback \cite{zheng2022}, spatial-bin occupancies \cite{bandyopadhyay2017}, or sample-point masses \cite{cai2026meanshift}.
PDE-constrained optimization solves an optimal control problem for the density dynamics, yielding boundary actuation \cite{sinigaglia2022} or velocity fields that need neither online density estimation nor communication \cite{sinigaglia2025}.
Both generate complex patterns but describe the target through a full density profile or its discretization.
Moment-based control is a finite-dimensional alternative when the task specifies selected distributional features: moments are invariant to agent labeling and can be estimated through neighbor communication without reconstructing a density.

Early moment-based designs regulate first and second moments.
Belta and Kumar \cite{belta2004} control a robot group through an abstraction given by its centroid, orientation, and covariance-based spread; Freeman et al.\ \cite{freeman2006} and Yang et al.\ \cite{yang2008} estimate the first and second inertial moments by consensus while the agents drive them to prescribed values, with convergence conditions for the coupled estimation and control dynamics; and our dispersion formation control \cite{chen2026} regulates the covariance eigenvalues while preserving the centroid.
Second-order descriptors, however, fix only the position, principal directions, and spreads.
Centering the positions, expressing them along the principal axes, and normalizing each spread yields a whitened cloud with zero mean and identity covariance that these descriptors do not see: very different shapes, such as a Gaussian-like cloud and the airplane of Fig.~\ref{fig:concept}, can share exactly the same covariance.

Higher-order descriptors originate in image analysis, where an image is represented by its moments with respect to polynomial bases, such as Legendre and Zernike polynomials, that are orthogonal over the image domain \cite{teague1980,teh1988}.
Liu et al.\ \cite{liu2024} transfer this representation to swarms: the Legendre or pseudo-Zernike moments of a target image are broadcast to the robots, and each robot follows the negative gradient of a weighted moment-error cost, estimated by a self-healing distributed estimator; this requires each robot to know its own absolute position in a common frame, provided by an external tracking system in the experiments, which achieve higher-order shapes with $50$ robots under heavy packet loss.
Alternatively, robots can be assigned to goal locations sampled from the target shape in a common frame, typically provided by external localization \cite{alonsomora2012,turpin2014,wang2020}, or self-assemble it through local rules \cite{rubenstein2014}; assignment reproduces the target support precisely but ties each robot to a specific goal, whereas moment specifications prescribe a class of configurations rather than a unique one.

Regulating higher moments through agent velocities raises three difficulties.
First, the orders are coupled: a velocity that corrects a third moment generally also changes the centroid and covariance.
The bases in \cite{liu2024} are orthogonal over the continuous image domain, not over the finite set of agent positions, so a gradient step toward one order perturbs the others, and the orders are balanced by empirically tuned gains.
Second, a finite swarm has finitely many velocity directions; once lower moments are preserved, a nonzero higher-order error may produce no corrective velocity, depending on the configuration and not only on the order (in the plane, for instance, full third-order regulation needs at least five agents).
Third, moments are global quantities, and estimating them while the formation moves couples estimation errors to the motion; in \cite{liu2024}, local stability is shown at full-rank equilibria with exact moments, and the coupled estimation and control dynamics are left for future work.

This paper addresses these difficulties while retaining the label-free moment specification.
We define desired formations by realizable central moments through any finite order in arbitrary spatial dimension, normalized by the covariance and aligned by one common orthogonal transformation.
Unlike moments evaluated in a fixed frame on a bounded domain, which prescribe absolute position, size, and orientation together with shape, this specification separates translation and principal spreads from higher-order shape and recovers the dispersion objective of \cite{chen2026} at order two.
We project negative-gradient velocity directions onto the orthogonal complements of polynomial sample spaces, that is, we orthogonalize with respect to the swarm's own empirical inner product rather than a fixed image domain.
The resulting hierarchical controller preserves every lower-order moment exactly under each higher-order correction, so adding orders never undoes the dispersion or lower-order shape already achieved; its error dynamics are lower triangular and converge locally exponentially for any positive gains.
A joint alternative regulates the selected higher moments together while preserving the centroid and covariance dynamics.

We give necessary bounds on the number of agents and sufficient polynomial-sampling conditions for regularity.
Both controllers admit distributed implementation through a two-layer estimator that needs only relative-position measurements and neighbor communication, with no global positioning system, and a slow--fast analysis shows that they retain local convergence under sufficiently fast and accurate estimation and a smooth principal-frame realization.
In a distributed simulation on a sparse communication graph, $350$ agents form an airplane from an unstructured swarm through moments up to order ten; the hierarchy reaches the full moment specification, whereas the joint law approaches it with smaller velocities.

Section~\ref{sec:problem} formulates the problem, Section~\ref{sec:general} develops the projected moment controllers, Section~\ref{sec:distributed} presents the distributed implementation, and Sections~\ref{sec:numerics} and~\ref{sec:conclusion} give numerical studies and conclusions.

\section{Preliminaries and Problem Formulation}\label{sec:problem}

\subsection{Notation}\label{sec:notation}
Let $N$ be the number of agents and $d$ the spatial dimension. Indices $i,j\in\{1,\ldots,N\}$ identify agents.
For a vector $x_i\in\mathbb R^d$, $(x_i)_{a_1}$ denotes its $a_1$-th coordinate, where $a_1\in\{1,\ldots,d\}$; the agent and coordinate indices are independent.
We write $x=(x_1,\ldots,x_N)$ for a collection of agent vectors.
Given one quantity $f_i$ per agent, $i=1,\ldots,N$, which may be a scalar, a vector, or a matrix, $\empavg{f}\coloneqq N^{-1}\sum_i f_i$ denotes its entrywise average over the agents, and $\empavg{f(x)}\coloneqq N^{-1}\sum_i f(x_i)$ for a function $f$ evaluated at the agent vectors; these are finite averages, not probabilistic expectations.
The symbols $\top$, $I_d$, $\mathbf1$, and $\|\cdot\|$ denote transpose, the $d\times d$ identity matrix, the all-ones vector of the required dimension, and the Euclidean vector norm (also written $\|\cdot\|_2$).
For matrices $A$ and $B$, $A\otimes B$ denotes their Kronecker product; for vectors, $\otimes$ denotes the tensor product of Section~\ref{sec:tensors}.
For matrices of the same size, $\langle A,B\rangle_F\coloneqq\operatorname{tr}(A^\top B)$ and $\|A\|_F^2\coloneqq\langle A,A\rangle_F$ are the Frobenius inner product and squared norm.
Here $\operatorname{tr}$ denotes the trace, and $\operatorname{diag}(x)$ forms a diagonal matrix from a vector $x$.
For a symmetric matrix, $A\succ0$ ($A\succeq0$) means positive (semi)definiteness.
The orthogonal group is $O(d)\coloneqq\{Q\in\mathbb R^{d\times d}:Q^\top Q=I_d\}$.

\subsection{Graph Theory}\label{sec:graph}
The communication network is an undirected graph $\mathcal G=(\mathcal V,\mathcal E)$, where $\mathcal V=\{1,\ldots,N\}$ is the agent set and $\mathcal E$ is the edge set, with no self-loops.
Agent $i$ has neighbors $\mathcal N_i\coloneqq\{j:(i,j)\in\mathcal E\}$.
The adjacency matrix $A=[a_{ij}]\in\mathbb R^{N\times N}$ has $a_{ij}=1$ on edges and zero otherwise; the Laplacian is $L_{\mathcal G}\coloneqq\operatorname{diag}(A\mathbf1)-A$.
It is symmetric positive semidefinite and satisfies $L_{\mathcal G}\mathbf1=0$.
If $\mathcal G$ is connected, its nullspace is $\ker L_{\mathcal G}=\operatorname{span}\{\mathbf1\}$ and its second-smallest eigenvalue satisfies $\lambda_2(L_{\mathcal G})>0$.
The distributed implementation in Section~\ref{sec:distributed} uses a fixed connected graph.
We also allow symmetric positive edge weights $a_{ij}=a_{ji}$, with the same Laplacian formula and properties.

\subsection{From matrices to tensors}\label{sec:tensors}
Each covariance entry averages a product of two position coordinates, or simply \emph{coordinates} for short, so the covariance needs two indices and is a matrix.
A third-order moment averages products of three coordinates and needs three indices; an order-$m$ moment needs $m$.
Tensors are the arrays that hold such data.
An order-$m$ tensor $T$ over $\mathbb R^d$ has entries $T_{a_1\cdots a_m}$, where each index $a_\ell\in\{1,\ldots,d\}$ selects a coordinate; the array has $d^m$ entries, and indices may repeat.
These coordinate indices are unrelated to the agent index $i$.
Vectors and matrices are tensors of orders one and two.

\emph{Tensor powers.}
The tensor product multiplies entries: for $x,y\in\mathbb R^d$, $(x\otimes y)_{a_1a_2}=(x)_{a_1}(y)_{a_2}$, so $x\otimes y=xy^\top$.
Repeating it gives the tensor power
\begin{equation*}
 (x^{\otimes m})_{a_1\cdots a_m}=\prod_{\ell=1}^m(x)_{a_\ell},
\end{equation*}
whose entries are all products of $m$ coordinates of $x$.
The average $\mathbb E_N$ applies entrywise to tensors as well. Thus, for the centered (barycentric) positions $z_i$ of Section~\ref{sec:moment-formation}, the covariance is $\empavg{zz^\top}=\empavg{z\otimes z}$, and the order-$m$ moment is $\empavg{z^{\otimes m}}$.
As a running example, take $d=2$, $m=3$, and $x=(1,2)^\top$.
Fixing the last index of $T\coloneqq x^{\otimes3}$ displays it as two $2\times2$ slices, $T_{::a_3}$ with entries $(T_{::a_3})_{a_1a_2}=T_{a_1a_2a_3}$:
\begin{equation*}
 T_{::1}=\begin{pmatrix}1&2\\2&4\end{pmatrix},\qquad
 T_{::2}=\begin{pmatrix}2&4\\4&8\end{pmatrix}.
\end{equation*}
Each entry equals $2^k$, where $k$ is the number of indices equal to $2$; e.g., $T_{122}=4$.

\emph{Symmetry.}
A tensor is symmetric if permuting its indices leaves its entries unchanged, as for $x^{\otimes m}$ and for every moment tensor.
An entry of a symmetric tensor depends only on how many times each coordinate appears among its indices, so the space $\operatorname{Sym}^m(\mathbb R^d)$ of symmetric tensors has $\binom{d+m-1}{m}$ independent entries.
For $d=2$ this count is $m+1$: in the example, $T_{111}=1$, $T_{112}=T_{121}=T_{211}=2$, $T_{122}=T_{212}=T_{221}=4$, and $T_{222}=8$.

\emph{Inner products and norms.}
For tensors $T,U$ of the same order, the Frobenius inner product and norm extend the matrix definitions:
\begin{equation*}
 \langle T,U\rangle_F=\sum_{a_1,\ldots,a_m=1}^d
 T_{a_1\cdots a_m}U_{a_1\cdots a_m},\quad
 \|T\|_F^2=\langle T,T\rangle_F.
\end{equation*}
All $d^m$ entries are counted, including repeated entries of symmetric tensors.
In the example, the entries $1,2,4,8$ occur $1,3,3,1$ times, so $\|T\|_F^2=1+3(4)+3(16)+64=125$, in agreement with $\|x^{\otimes m}\|_F^2=\|x\|^{2m}=5^3$.

\emph{Contractions.}
Contracting an order-$m$ tensor $D$ with $m-1$ copies of $x$ sums over all but its first index and leaves a vector $D(x^{\otimes(m-1)})\in\mathbb R^d$:
\begin{equation*}
 \big(D(x^{\otimes(m-1)})\big)_{a_1}=\sum_{a_2,\ldots,a_m=1}^d D_{a_1\cdots a_m}\prod_{\ell=2}^m(x)_{a_\ell}.
\end{equation*}
At order two this is the matrix--vector product $D(x)=Dx$; in the example, $T(x^{\otimes2})=x(x^\top x)^2=(25,50)^\top$.
Contractions produce velocities from moment errors.
For a fixed symmetric $D$, the scalar $\langle D,x^{\otimes m}\rangle_F$ is a homogeneous polynomial of degree $m$ in $x$, with gradient
\begin{equation}\label{eq:tensor-gradient}
 \nabla_x\langle D,x^{\otimes m}\rangle_F=m\,D(x^{\otimes(m-1)}),
\end{equation}
which generalizes $\nabla_x(x^\top Dx)=2Dx$.
Hence the average $\empavg{\langle D,z^{\otimes m}\rangle_F}=\langle D,\empavg{z^{\otimes m}}\rangle_F$ has partial derivative $(m/N)\,D(z_i^{\otimes(m-1)})$ with respect to $z_i$, and the controllers in Section~\ref{sec:general} are built from such contractions.

\emph{Coordinate transformations.}
For $A\in\mathbb R^{d\times d}$, $A^{\otimes m}T$ applies $A$ to every index of $T$:
\begin{equation*}
 (A^{\otimes m}T)_{a_1\cdots a_m}=\sum_{b_1,\ldots,b_m=1}^d A_{a_1b_1}\cdots A_{a_mb_m}T_{b_1\cdots b_m}.
\end{equation*}
At order two this is $ATA^\top$, and $(Ax)^{\otimes m}=A^{\otimes m}x^{\otimes m}$, so transforming all positions by $A$ transforms every moment tensor by $A^{\otimes m}$.
For example, $A=\operatorname{diag}(2,1)$ doubles the first coordinate, and each index equal to $1$ contributes a factor $2$: the example entries $T_{111},T_{112},T_{122},T_{222}$ are multiplied by $8,4,2,1$.
Two cases are used later: orthogonal $A=Q\in O(d)$, a common rotation or reflection that aligns moments in Definition~\ref{def:moment-formation}, and the non-orthogonal whitening map of Section~\ref{sec:whitening}, which normalizes the principal spreads.

\subsection{From dispersion formation to arbitrary-moment formation}\label{sec:dispersion}\label{sec:moment-formation}
\begin{figure*}[!t]
 \centering
 \includegraphics[width=\textwidth]{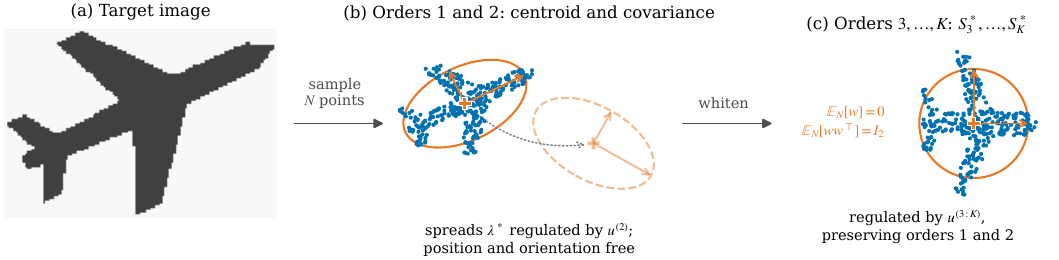}
 \caption{From a target image to moment specifications. (a) Binary target image. (b) Reference configuration of $N=350$ agents sampled from the image support, with its centroid (order one) and covariance (order two; ellipse at two standard deviations with principal axes). The dispersion controller $u^{(2)}$ regulates the principal spreads $\lambda^*$; without a prescribed centroid ($u^{(1)}=0$), the position and orientation of the formation remain free, and the dashed ellipse is an equally valid placement. (c) Whitened reference $w^*$ with zero mean and identity covariance; its standardized moments $S_3^*,\ldots,S_K^*$ encode the remaining shape and are regulated by $u^{(3:K)}$ without altering orders one and two.}
 \label{fig:concept}
\end{figure*}
Consider a configuration $p=(p_1,\ldots,p_N)\in(\mathbb R^d)^N$.
Its centroid, centered positions, and covariance are
\begin{equation}\label{eq:covariance}
 p_c\coloneqq\empavg{p},\quad z_i\coloneqq p_i-p_c,\quad
 C(p)\coloneqq\empavg{zz^\top}.
\end{equation}
Given a reference configuration $p^*$, define $p_c^*$ and $C^*\coloneqq C(p^*)$ in the same way. Covariance eigenvalues measure the swarm's variances along its principal directions. Dispersion formation control \cite{chen2026} regulates these collective spreads to prescribed values.

\begin{defn}[Covariance similarity and desired dispersion]\label{def:dispersion}
Two configurations are covariance similar if their covariance matrices have the same eigenvalues. Since these matrices are symmetric, the desired dispersion set is
\begin{equation}\label{eq:dispersion-set}
 \mathcal F_2\coloneqq\{p:\ C^*=QC(p)Q^\top\text{ for some }Q\in O(d)\}.
\end{equation}
\end{defn}

Here $Q$ aligns the covariances, leaving translation and orthogonal orientation free. Distinct shapes may share this dispersion; higher moments capture additional spatial features.

\smallskip\noindent\emph{Higher-order formation descriptors.}
Generalizing the covariance from products of two centered coordinates to products of $m$, define the order-$m$ central moment for $m\ge2$ by
\begin{equation}\label{eq:central-moments}
 M_m(p)\coloneqq\empavg{z^{\otimes m}},\quad
 (M_m)_{a_1\cdots a_m}=\frac1N\sum_i\prod_{\ell=1}^m(z_i)_{a_\ell}.
\end{equation}
Thus $M_m(p)\in\operatorname{Sym}^m(\mathbb R^d)$, and $M_2=C$ recovers the covariance. The empirical average acts componentwise on the tensor entries.

For example, for $d=2$, $M_3$ has four independent entries:
\begin{equation*}
\begin{aligned}
 (M_3)_{111}&=\empavg{(z_i)_1^3},&(M_3)_{112}&=\empavg{(z_i)_1^2(z_i)_2},\\
 (M_3)_{122}&=\empavg{(z_i)_1(z_i)_2^2},&(M_3)_{222}&=\empavg{(z_i)_2^3}.
\end{aligned}
\end{equation*}
The remaining entries follow by symmetry, e.g., $(M_3)_{112}=(M_3)_{121}=(M_3)_{211}$.
These quantities can be stored as a vector; the tensor representation makes their symmetry and transformation under a change of coordinates explicit: an orthogonal $Q$ maps $M_m$ to $Q^{\otimes m}M_m$ (Section~\ref{sec:tensors}).

Let $K\ge2$ be the largest prescribed moment order, and let $M_m^*\coloneqq M_m(p^*)$, $m=2,\ldots,K$, be desired moments of a reference configuration with $C^*\succ0$.
Using a reference of the same $N$ agents ensures that the entire desired sequence is realizable.
Alternatively, the tensors may be prescribed directly, provided they are jointly realizable by $N$ equal-weight points.
We extend covariance similarity by requiring one common orthogonal transformation to align all central moments through order $K$.

\begin{defn}[Moment similarity through order $K$]\label{def:moment-formation}
For $K\ge2$, two configurations $p$ and $p^*$ are moment similar through order $K$ if one common $Q\in O(d)$ maps each $M_m(p)$ to $M_m(p^*)$, $m=2,\ldots,K$.
Given the reference moments $\mathcal M_K^*\coloneqq(M_2^*,\ldots,M_K^*)$, the corresponding moment-similarity class is the desired formation set
\begin{equation}\label{eq:moment-set}
\begin{split}
 \mathcal F_K\coloneqq\{p:\ &\exists Q\in O(d)\text{ such that}\\
 &Q^{\otimes m}M_m(p)=M_m^*,\quad m=2,\ldots,K\}.
\end{split}
\end{equation}
\end{defn}

For $K=2$, moment similarity reduces to covariance similarity in Definition~\ref{def:dispersion}.
For larger $K$, it retains the dispersion requirement and adds higher-order shape descriptors, all expressed relative to the same reference orientation.
Central moments leave translation free, and the specification does not assign individual agents to reference points.
Finite-order moment similarity may relate geometrically different configurations; it does not require a pointwise rotation, reflection, translation, or scaling between them.
Fig.~\ref{fig:concept} illustrates the resulting design pipeline.
A target image is sampled into a reference configuration, whose covariance prescribes the principal spreads while leaving the position and orientation of the formation free; its higher-order moments, expressed in the whitened coordinates of Section~\ref{sec:whitening}, encode the remaining shape.

\subsection{Control objective}\label{sec:control-problem}
Consider the single-integrator agent dynamics
\begin{equation}\label{eq:agent-dynamics}
 \dot p_i=u_i,\quad i=1,\ldots,N,
\end{equation}
where $u_i\in\mathbb R^d$ is the control input of agent $i$.
In compact form, $\dot p=u$ with the stacked input $u\coloneqq(u_1,\ldots,u_N)\in(\mathbb R^d)^N$.
Single integrators keep the focus on the moment-control mechanism: $u_i$ serves as a guidance velocity that a lower-level controller can track for vehicles with more complex dynamics.
For a prescribed admissible moment tuple $\mathcal M_K^*$, define the formation error
\begin{equation}\label{eq:formation-error}
 e_K(p)\coloneqq\min_{Q\in O(d)}
 \left(\sum_{m=2}^K\|Q^{\otimes m}M_m(p)-M_m^*\|_F^2\right)^{1/2}.
\end{equation}
The minimum is attained because $O(d)$ is compact, and $e_K(p)=0$ exactly when $p\in\mathcal F_K$.

\begin{prob}[Arbitrary-moment formation control]\label{prob:control}
For \eqref{eq:agent-dynamics} and desired moments $\mathcal M_K^*$ with $M_2^*\succ0$, design a feedback law $u$ and establish conditions on the reference and initial configuration under which
\begin{equation}\label{eq:control-objective}
 \lim_{t\to\infty}e_K(p(t))=0.
\end{equation}
That is, the formation must approach the reference moment-similarity class $\mathcal F_K$ through order $K$.
\end{prob}

If a desired centroid $p_c^*$ is also specified, the additional objective is $p_c(t)\to p_c^*$.
The following sections develop feedback constructions and convergence conditions for general $K$, illustrated at order three.

\section{Moment Control by Polynomial Projection}\label{sec:general}
This section designs centralized controllers that assume exact access to the current configuration statistics; Section~\ref{sec:distributed} removes this assumption.
We separate the centroid and covariance from the higher-order shape through whitened coordinates, construct projected gradient feedback for an arbitrary moment order, illustrate it at order three, and then analyze the hierarchical and joint controllers.

\subsection{Whitened coordinates and moment errors}\label{sec:whitening}
We use the centroid as the position origin and covariance principal directions and scales to define whitened coordinates. The dispersion controller regulates these scales, while the higher-moment controller reshapes the whitened cloud without altering the centroid and covariance dynamics.
Choose eigendecompositions $C^*=V^*\Lambda^*(V^*)^\top$ and $C=V\Lambda V^\top$, where $V^*,V\in O(d)$ contain unit eigenvectors and $\Lambda^*=\operatorname{diag}(\lambda^*)\succ0$, $\Lambda=\operatorname{diag}(\lambda)\succ0$ contain the corresponding eigenvalues.
The feedback
\begin{equation}\label{eq:base}
 u_i^{(2)}\coloneqq-V\operatorname{diag}(\lambda-\lambda^*)V^\top z_i
\end{equation}
adjusts the spread along each principal axis toward its desired value \cite{chen2026}.
Here $V^\top z_i$ gives the agent's displacements along the principal axes: the diagonal gain contracts a displacement when the corresponding variance exceeds its target and expands it when the variance is below target.
Multiplication by $V$ maps these axiswise velocities back to physical coordinates.
It preserves the centroid and principal directions, with
$\frac{d}{dt}(\lambda)_{a_1}=-2(\lambda)_{a_1}((\lambda)_{a_1}-(\lambda^*)_{a_1})$.
Assume $(\lambda(0))_{a_1},(\lambda^*)_{a_1}>0$.
Choose $V$ at $t=0$ and retain its labels, including at coincident eigenvalues.

To separate these axis scales from higher-order distribution features, define
\begin{equation}\label{eq:white}
 w_i\coloneqq\Lambda^{-1/2}V^\top z_i,\quad S_m\coloneqq\empavg{w^{\otimes m}},
\end{equation}
where $(w_i)_{a_1}=(V^\top z_i)_{a_1}/\sqrt{(\lambda)_{a_1}}$. The $a_1$th column of $V$ is the unit eigenvector of $C$ corresponding to $(\lambda)_{a_1}$.
Thus $w_i$ is the centered position expressed along the principal axes, with each displacement measured in units of its standard deviation $\sqrt{(\lambda)_{a_1}}$.
This normalization, known as PCA whitening \cite{kessy2018} and used to normalize image moments \cite{flusser2016} (Fig.~\ref{fig:concept}(c)), gives
\begin{equation*}
 \empavg{w}=0,\quad \empavg{ww^\top}
 =\Lambda^{-1/2}V^\top CV\Lambda^{-1/2}=I_d.
\end{equation*}
Consequently, $S_2=I_d$, and the dimensionless tensors $S_m$, $m\ge3$, describe features beyond covariance, such as asymmetry.
For example, in one dimension $S_3=M_3/\lambda^{3/2}$ is skewness.
Whitening is a coordinate transformation: the physical covariance need not be the identity.
The higher-moment errors can then compare normalized distribution features without mixing them with changes in the principal-axis spreads.

Under \eqref{eq:base}, each principal-axis displacement and its standard deviation scale together, yielding
\begin{equation}\label{eq:white-invariance}
 \frac{d}{dt}(w_i)_{a_1}
 =-((\lambda)_{a_1}-(\lambda^*)_{a_1})(w_i)_{a_1}
 -\frac{(\dot\lambda)_{a_1}}{2(\lambda)_{a_1}}(w_i)_{a_1}=0.
\end{equation}
Hence dispersion feedback leaves the normalized point cloud unchanged.
The added feedback will reshape this cloud while preserving its zero mean and identity covariance, allowing higher moments to be controlled independently of the axis scales.

Normalize the reference in the same way, and define the target tensors $S_m^*$ and error tensors $E_m\in\operatorname{Sym}^m(\mathbb R^d)$:
\begin{align}\label{eq:targets}
 w_i^*&\coloneqq(\Lambda^*)^{-1/2}(V^*)^\top(p_i^*-p_c^*),\nonumber\\
 S_m^*&\coloneqq\empavg{(w^*)^{\otimes m}},\quad E_m\coloneqq S_m-S_m^*,
 \quad 3\le m\le K.
\end{align}
Equivalently,
\begin{equation*}
 S_m^*=\big[(\Lambda^*)^{-1/2}(V^*)^\top\big]^{\otimes m}M_m^*.
\end{equation*}
The inverse relation $z_i=V\Lambda^{1/2}w_i$ restores orientation and scale, so no moment information is lost:
\begin{equation}\label{eq:moment-reconstruction}
 M_m(p)=(V\Lambda^{1/2})^{\otimes m}S_m.
\end{equation}
With $Q\coloneqq V^*V^\top$, $\lambda\to\lambda^*$ and $E_m\to0$ imply
\begin{align*}
 Q^{\otimes m}M_m(p)&=(V^*\Lambda^{1/2})^{\otimes m}S_m\\
 &\longrightarrow
 \big[V^*(\Lambda^*)^{1/2}\big]^{\otimes m}S_m^*=M_m^*.
\end{align*}
Thus these two control tasks achieve \eqref{eq:control-objective} with one common alignment for all orders.
If a desired centroid $p_c^*$ is specified, the optional centroid controller is
\begin{equation}\label{eq:centroid-control}
 u_i^{(1)}\coloneqq-k_c(p_c-p_c^*),\quad k_c>0.
\end{equation}
Otherwise, set $u_i^{(1)}\coloneqq0$.

\subsection{Projected gradient feedback}\label{sec:projected}
We project a negative-gradient candidate for each moment order onto a velocity subspace that preserves all lower-order moments of the whitened cloud.
For a vector-valued field $v$, write $v_i=v(w_i)\in\mathbb R^d$ for its value at agent $i$, and identify its sampled values with $v=(v_1,\ldots,v_N)$ when taking empirical inner products.
For two sampled fields $v,h$, set
\begin{equation*}
 \langle v,h\rangle_N\coloneqq\frac1N\sum_i v_i^\top h_i,
 \quad \|v\|_N^2\coloneqq\langle v,v\rangle_N.
\end{equation*}
Dependence on the current configuration is suppressed.
We denote the order-$m$ corrective field by $v^{(m)}$; its value at agent $i$ is $v^{(m)}(w_i)$, and its $a_1$th coordinate is $(v^{(m)}(w_i))_{a_1}$.
The superscript identifies the controlled moment order, while each field value remains a vector in $\mathbb R^d$.

For $3\le m\le K$, consider the error energy $J_m(w)\coloneqq\tfrac12\|E_m\|_F^2$.
Since $S_m=\empavg{w^{\otimes m}}$ and $E_m$ is symmetric, the gradient identity \eqref{eq:tensor-gradient} gives
\begin{equation}\label{eq:gradient}
 \nabla_{w_i}J_m=\frac mN\,g^{(m)}(w_i),\quad
 g^{(m)}(w)\coloneqq E_m(w^{\otimes(m-1)}).
\end{equation}
Thus $mg^{(m)}$ is the gradient of $J_m$ under the empirical inner product, and $-k_mg^{(m)}$, with $k_m>0$, is a negative-gradient candidate.
This candidate, however, generally also changes the centroid, the covariance, and the lower-order moments.
For instance, $\frac{d}{dt}\empavg{w}=\empavg{\dot w}$ and $\frac{d}{dt}\empavg{ww^\top}=\empavg{\dot ww^\top+w\dot w^\top}$, so a velocity field preserves the mean and covariance of the whitened cloud when each of its components is empirically orthogonal to the constant and to every coordinate.
The following lemma extends this observation to all lower orders.
\begin{lem}[Freezing lower moments]\label{lem:freeze}
Let $r\ge1$ be an integer and every component of a common agent velocity field be empirically orthogonal to every polynomial of degree at most $r-1$.
Then its contribution to the derivative of every raw moment of order at most $r$ vanishes.
The same holds for central moments.
\end{lem}
\begin{pf}
Write $v_i\coloneqq v(p_i)\in\mathbb R^d$ for the common velocity field evaluated at agent $i$, with $(v_i)_{a_1}$ its $a_1$th coordinate component.
Dependence of $v$ on time and on the current configuration is suppressed.
The empirical orthogonality assumption means
\begin{equation}\label{eq:freeze-orthogonality}
 \frac1N\sum_{i=1}^N (v_i)_{a_1} q(p_i)=0,
 \quad a_1=1,\ldots,d,\quad \deg q\le r-1,
\end{equation}
where $q:\mathbb R^d\to\mathbb R$ is any scalar polynomial satisfying this degree bound and $\deg q$ denotes its total degree.
To compute this field's contribution, set $\dot p_i=v_i$ and denote the order-$s$ raw moment by $\Gamma_s\coloneqq\empavg{p^{\otimes s}}$.
Fix $1\le s\le r$ and coordinate indices $a_1,\ldots,a_s\in\{1,\ldots,d\}$.
By the product rule, the corresponding raw-moment component satisfies
\begin{equation*}
 \frac{d}{dt}(\Gamma_s)_{a_1\cdots a_s}
 =\sum_{k=1}^s\frac1N\sum_{i=1}^N
 (v_i)_{a_k}\prod_{\ell\ne k}(p_i)_{a_\ell}.
\end{equation*}
For each $k$, the remaining product is $q_k(p_i)\coloneqq\prod_{\ell\ne k}(p_i)_{a_\ell}$, where $q_k:\mathbb R^d\to\mathbb R$ is a scalar monomial.
Thus the $k$th inner average is exactly $N^{-1}\sum_i (v_i)_{a_k}q_k(p_i)$.
This monomial has total degree $s-1\le r-1$; when $s=1$, the empty product is $q_1=1$.
Thus every inner average vanishes by \eqref{eq:freeze-orthogonality}, proving $\dot\Gamma_s=0$.
In particular, $\Gamma_1=p_c$ is fixed by this velocity contribution.
Expanding the factors in $\empavg{(p-p_c)^{\otimes s}}$ expresses each central-moment component as a polynomial in raw-moment components of orders at most $s$.
Their derivatives are all zero, so the central-moment derivatives also vanish.
\end{pf}

Lemma~\ref{lem:freeze} suggests removing from the candidate its components along low-degree polynomials.
For any integer $r\ge0$, let $\mathcal P_r\subset\mathbb R^N$ be the sample space of polynomials of total degree at most $r$. Let $q_r(w)\in\mathbb R^{D_r}$ list all distinct monomials of these degrees, including $1$, where $D_r\coloneqq\binom{d+r}{r}$.
Define $Q_r\in\mathbb R^{N\times D_r}$ by row $i$ equal to $q_r(w_i)^\top$; thus $\mathcal P_r$ is its column space. For $m\ge3$, use $r=m-2$.
On a neighborhood where $Q_{m-2}$ has full column rank, the empirical residual projector is
\begin{equation}\label{eq:project}
 R_{m-2}\coloneqq I_N-Q_{m-2}(Q_{m-2}^\top Q_{m-2})^{-1}Q_{m-2}^\top.
\end{equation}
Here $Q_{m-2}$ spans $\mathcal P_{m-2}$, so the subtracted term is $\operatorname{proj}_{\mathcal P_{m-2}}$ and $R_{m-2}$ projects onto $\mathcal P_{m-2}^\perp$.
It acts separately on each spatial component: for $y\in\mathbb R^N$, $R_{m-2}y$ is the vector closest to $y$ among all $y'\in\mathbb R^N$ with $Q_{m-2}^\top y'=0$, that is, among vectors empirically orthogonal to $\mathcal P_{m-2}$.
An implementation should use an orthogonal factorization rather than form the inverse explicitly.

In the next formula, brackets arrange the sampled vectors as rows of an $N\times d$ array, so $R_{m-2}$ acts on each coordinate column. For $D\in\operatorname{Sym}^m(\mathbb R^d)$ set
\begin{equation}\label{eq:Am}
 \mathcal A_mD\coloneqq R_{m-2}\big[D(w_i^{\otimes(m-1)})\big]_{i=1}^N,
 \quad \mathcal H_m\coloneqq\mathcal A_m^*\mathcal A_m.
\end{equation}
Here $\mathcal A_m^*$ is the adjoint of $\mathcal A_m$, defined by $\langle\mathcal A_mD,v\rangle_N=\langle D,\mathcal A_m^*v\rangle_F$ for every $D\in\operatorname{Sym}^m(\mathbb R^d)$ and sampled velocity field $v$, so $\mathcal H_m$ is self-adjoint and positive semidefinite on $\operatorname{Sym}^m(\mathbb R^d)$.
For such an operator, $\mathcal H\succ0$ means $\langle D,\mathcal H D\rangle_F>0$ for every nonzero $D$, and $\mathcal H\succeq\gamma I$ means $\langle D,\mathcal H D\rangle_F\ge\gamma\|D\|_F^2$ for all $D$, with $I$ the identity operator.
For gains $k_m>0$, the hierarchical feedback and its physical realization are
\begin{align}\label{eq:hierarchy}
 v^{(m)}&\coloneqq-k_m\mathcal A_m E_m,\quad
 \dot w=\sum_{m=3}^K v^{(m)},\\
 u_i&\coloneqq u_i^{(1)}+u_i^{(2)}+u_i^{(3:K)},\label{eq:physical}
\end{align}
where
\begin{equation}\label{eq:higher-physical}
 \begin{aligned}
 u_i^{(3:K)}&\coloneqq\sum_{m=3}^K u_i^{(m)}
 =V\Lambda^{1/2}\sum_{m=3}^K v^{(m)}(w_i),\\
 u_i^{(m)}&\coloneqq V\Lambda^{1/2}v^{(m)}(w_i),\quad 3\le m\le K.
 \end{aligned}
\end{equation}
By \eqref{eq:gradient}, $\mathcal A_mE_m=R_{m-2}g^{(m)}$, so $v^{(m)}$ is the velocity field in $\mathcal P_{m-2}^\perp$ closest to the candidate $-k_mg^{(m)}$; when nonzero, its normalization gives steepest descent of $J_m$ among unit-norm velocity fields in this subspace.
The term $u_i^{(3:K)}$ combines the controllers for moments of orders $3$ through $K$ while preserving the centroid and covariance dynamics.
Each residual has a polynomial representative of degree at most $m-1$.
This is a Gram--Schmidt construction in the swarm's empirical inner product; orthogonal-polynomial recurrences provide an alternative basis construction \cite{xu2004}.
Image-moment bases are orthogonal over a fixed continuous domain \cite{teague1980,teh1988,liu2024}; here orthogonality is taken over the current agent positions, which is what makes the preservation of lower-order moments exact for a finite swarm.

\subsection{Third-order example}\label{sec:third}
For $m=3$, the projection has a closed form.
The space $\mathcal P_1$ is spanned by the constant sample vector $\mathbf1$ and the coordinate sample vectors $\chi^{(a_1)}\coloneqq((w_1)_{a_1},\ldots,(w_N)_{a_1})^\top$, $a_1=1,\ldots,d$, and whitening makes this basis empirically orthonormal:
\begin{equation*}
 \langle\mathbf1,\mathbf1\rangle_N=1,\quad
 \langle\mathbf1,\chi^{(a_1)}\rangle_N=0,\quad
 \langle\chi^{(a_1)},\chi^{(a_2)}\rangle_N=\delta_{a_1a_2},
\end{equation*}
where $\delta_{a_1a_2}$ equals one if $a_1=a_2$ and zero otherwise.
Projecting the quadratic samples $((w_i)_{a_2}(w_i)_{a_3})_{i=1}^N$ onto this basis therefore requires only $\empavg{(w_i)_{a_2}(w_i)_{a_3}}=\delta_{a_2a_3}$ and $\empavg{(w_i)_{a_2}(w_i)_{a_3}(w_i)_{a_4}}=(S_3)_{a_2a_3a_4}$, and leaves the residual
\begin{equation}\label{eq:psi}
 \psi_{a_2a_3}(w)\coloneqq(w)_{a_2}(w)_{a_3}-\delta_{a_2a_3}-\sum_{a_4=1}^d(S_3)_{a_2a_3a_4}(w)_{a_4}.
\end{equation}
Since $g^{(3)}$ is linear in these quadratic products, the third-order feedback is
\begin{equation}\label{eq:A3}
 \big(v^{(3)}(w)\big)_{a_1}=-k_3\sum_{a_2,a_3=1}^d(E_3)_{a_1a_2a_3}\psi_{a_2a_3}(w).
\end{equation}
The quadratic term of $\psi_{a_2a_3}$ is the raw third-moment direction, and its constant and linear terms cancel the centroid and covariance changes that this direction would induce.
Evaluating \eqref{eq:A3} requires only moments through order three.
In one dimension, \eqref{eq:A3} reduces to $v^{(3)}(w)=-k_3E_3(w^2-1-S_3w)$, where $S_3$ is the skewness.
If the skewness is below its target, so that $E_3<0$, agents far from the centroid on either side move in the positive direction, whereas central agents move in the negative direction.
The right tail extends and the left tail contracts, which increases the skewness while the mean and variance remain fixed.

\subsection{Error dynamics and regularity}
The next result is the central structural property of the construction: corrections of higher order never disturb the moments of lower order.
A correction of order $r$ may, however, still change moments of higher order $m>r$.
By the product rule, the correction $v^{(r)}=-k_r\mathcal A_rE_r$ in \eqref{eq:hierarchy} changes $E_m$ by $\mathcal B_{mr}E_r$, where the coupling operator
\begin{equation}\label{eq:coupling-components}
 \begin{aligned}
 &(\mathcal B_{mr}(w)E_r)_{a_1\cdots a_m}\\
 &\quad\coloneqq-k_r\sum_{\ell=1}^m\empavg{\big((\mathcal A_rE_r)(w_i)\big)_{a_\ell}
 \prod_{\ell'\ne\ell}(w_i)_{a_{\ell'}}}
 \end{aligned}
\end{equation}
is linear in $E_r$ for fixed $w$.
\begin{thm}[Preservation of lower-order moments]\label{thm:hierarchy}
Suppose the sample matrix $Q_{m-2}$ in \eqref{eq:project} has full column rank for $3\le m\le K$. Under the controller \eqref{eq:physical}:
\begin{itemize}
\item[(i)] each correction $u^{(m)}$ in \eqref{eq:higher-physical}, $3\le m\le K$, does not affect any moment of order below $m$; in particular, the centroid and covariance follow the dynamics of $u^{(1)}$ in \eqref{eq:centroid-control} and $u^{(2)}$ in \eqref{eq:base} alone;
\item[(ii)] the moment errors $E_m$ in \eqref{eq:targets} satisfy the lower triangular system \eqref{eq:cascade}, where $\mathcal H_m$ is defined in \eqref{eq:Am} and the coupling operators $\mathcal B_{mr}$ in \eqref{eq:coupling-components} depend smoothly on $w$:
\end{itemize}
\begin{equation}\label{eq:cascade}
 \left(\begin{smallmatrix}\dot E_3\\[1pt]\dot E_4\\\vdots\\\dot E_K\end{smallmatrix}\right)
 =\left(\begin{smallmatrix}
 -3k_3\mathcal H_3 & 0 & \cdots & 0\\[1pt]
 \mathcal B_{43} & -4k_4\mathcal H_4 & \cdots & 0\\
 \vdots & \vdots & \ddots & \vdots\\
 \mathcal B_{K3} & \mathcal B_{K4} & \cdots & -Kk_K\mathcal H_K
 \end{smallmatrix}\right)
 \left(\begin{smallmatrix}E_3\\[1pt]E_4\\\vdots\\E_K\end{smallmatrix}\right)
\end{equation}
\end{thm}
\begin{pf}
By \eqref{eq:project}--\eqref{eq:hierarchy}, each component of $v^{(r)}$ is orthogonal to $\mathcal P_{r-2}$.
Lemma~\ref{lem:freeze}, with its parameter equal to $r-1$, therefore shows that $v^{(r)}$ preserves every moment of order below $r$.
In particular, all these fields preserve mean and covariance, so their sum leaves the physical centroid and covariance dynamics unchanged.
Differentiating \eqref{eq:white} as in \eqref{eq:white-invariance}, the contributions of $u^{(1)},u^{(2)}$ cancel, which gives $\dot w=\sum_{r=3}^K v^{(r)}$.
Since the targets are fixed, $\dot E_m=\dot S_m$.
The product rule gives
\begin{equation*}
 (\dot E_m)_{a_1\cdots a_m}
 =\sum_{r=3}^K\sum_{\ell=1}^m
 \empavg{(v^{(r)}(w_i))_{a_\ell}
 \prod_{\ell'\ne\ell}(w_i)_{a_{\ell'}}}.
\end{equation*}
For any symmetric order-$m$ tensor $D$, contraction makes the $m$ product-rule terms equal, yielding
\begin{equation}\label{eq:momentderivative}
 \langle D,\dot S_m\rangle_F
 =m\empavg{\dot w^\top D(w^{\otimes(m-1)})}.
\end{equation}
We separate the contributions according to $r$.
If $r>m$, the coordinate product has degree $m-1\le r-2$, so orthogonality to $\mathcal P_{r-2}$ makes each contribution zero.
For $r=m$, the difference between $D(w^{\otimes(m-1)})$ and $\mathcal A_mD$ lies componentwise in $\mathcal P_{m-2}$ and is orthogonal to $v^{(m)}$.
Thus the own-order contribution satisfies
\begin{equation*}
 \begin{aligned}
 m\langle v^{(m)},D(w^{\otimes(m-1)})\rangle_N
 &=m\langle v^{(m)},\mathcal A_mD\rangle_N\\
 &=-mk_m\langle\mathcal A_mE_m,\mathcal A_mD\rangle_N\\
 &=-mk_m\langle D,\mathcal H_mE_m\rangle_F.
 \end{aligned}
\end{equation*}
Since this holds for every symmetric $D$, that contribution is $-mk_m\mathcal H_mE_m$.
For $r<m$, substituting $v^{(r)}=-k_r\mathcal A_rE_r$ in the componentwise product rule gives exactly $\mathcal B_{mr}E_r$ in \eqref{eq:coupling-components}, whose coefficients are smooth in $w$ because the monomial evaluations and $(Q_{r-2}^\top Q_{r-2})^{-1}$ are smooth on the full-column-rank domain.
Stacking the contributions for $m=3,\ldots,K$ gives the lower triangular block system \eqref{eq:cascade}.
\end{pf}

Theorem~\ref{thm:hierarchy} means that the order-$m$ error is driven only by itself and by errors of lower order.
Once $E_3,\ldots,E_{m-1}$ vanish, $\dot E_m=-mk_m\mathcal H_mE_m$; in particular, when only $v^{(m)}$ is active,
\begin{equation}\label{eq:menergy}
 \frac{d}{dt}\tfrac12\|E_m\|_F^2
 =-mk_m\langle E_m,\mathcal H_mE_m\rangle_F
 =-mk_m\|\mathcal A_mE_m\|_N^2.
\end{equation}
Adding a higher order therefore never undoes the dispersion or the lower-order shape already achieved, and the gains need not weight the orders against each other.
The coupling operators may involve moments above order $K$, so $E_3,\ldots,E_K$ alone generally do not determine their derivatives.

By \eqref{eq:menergy}, a single active level dissipates every nonzero error $E_m$ exactly when $\mathcal H_m\succ0$, equivalently when $\mathcal A_m$ is injective: no nonzero order-$m$ error produces a zero corrective field.
We call this full order-$m$ regularity.
It requires enough agents and a suitably spread configuration.

\begin{prop}[Number of agents and regularity]\label{prop:agents}
Let $3\le m\le K$ and $h_m\coloneqq\binom{d+m-1}{m}$, the dimension of $\operatorname{Sym}^m(\mathbb R^d)$.
\begin{itemize}
\item[(i)] If $Q_{m-2}$ has full column rank and $\mathcal H_m\succ0$, then
\begin{equation}\label{eq:agent-number-bound}
 N\ge D_{m-2}+\left\lceil\frac{h_m}{d}\right\rceil,
\end{equation}
where $\lceil s\rceil$ is the smallest integer not less than $s$.
\item[(ii)] If $Q_{K-1}$ has full column rank, which requires $N\ge D_{K-1}$, then $Q_{m-2}$ has full column rank and $\mathcal H_m\succ0$ for every $3\le m\le K$.
\end{itemize}
\end{prop}
\begin{pf}
(i) Under the rank condition, $\mathcal P_{m-2}^\perp$ has dimension $N-D_{m-2}$, and each of the $d$ output components of $\mathcal A_m$ lies in it, so $\operatorname{rank}\mathcal A_m\le d(N-D_{m-2})$.
Injectivity on $\operatorname{Sym}^m(\mathbb R^d)$ requires $d(N-D_{m-2})\ge h_m$, which gives \eqref{eq:agent-number-bound} because $N$ is an integer.
(ii) The columns of $Q_{m-2}$ are among those of $Q_{K-1}$, hence linearly independent.
If $\mathcal A_mD=0$, each component of $D(w^{\otimes(m-1)})$ coincides on all agents with a polynomial of degree at most $m-2$.
Their difference is a polynomial of degree at most $m-1\le K-1$ that vanishes at every agent, hence vanishes identically by the rank condition.
Its homogeneous part of degree $m-1$, namely $D(w^{\otimes(m-1)})$, is therefore identically zero, and so is $\langle D,w^{\otimes m}\rangle_F=w^\top D(w^{\otimes(m-1)})$.
A symmetric tensor is determined by this polynomial, so $D=0$.
\end{pf}

For $m=3$, \eqref{eq:agent-number-bound} reads $N\ge d+1+\lceil(d+1)(d+2)/6\rceil$: at least five agents in the plane and eight in 3D space.
Condition (i) is necessary, whereas (ii) is sufficient and can be conservative.
For $d=2$ and $K=3$, (ii) requires six agents, but five suffice.
The whitened regular pentagon $w_j\coloneqq\sqrt2(\cos(2\pi j/5),\sin(2\pi j/5))$, $j=1,\ldots,5$, has $S_3=0$.
For $D\in\operatorname{Sym}^3(\mathbb R^2)$ with independent entries $(t_1,t_2,t_3,t_4)\coloneqq(D_{111},D_{112},D_{122},D_{222})$, discrete angular orthogonality gives
\begin{equation}\label{eq:pentagon}
 \|\mathcal A_3D\|_N^2
 =\tfrac12\big[(t_1-t_3)^2+(t_2-t_4)^2+4t_2^2+4t_3^2\big].
\end{equation}
The generalized eigenvalues relative to $\|D\|_F^2=t_1^2+3t_2^2+3t_3^2+t_4^2$ are $1/3,1/3,1,1$, so $\mathcal H_3\succeq\frac13I$ and the bound in (i) is attained.

\subsection{Local convergence of hierarchical control}\label{sec:stability}
We prove local exponential convergence despite the triangular couplings. The physical target is $\mathcal F_K$; in whitened coordinates, the aligned target set $\mathcal Z$ below matches the reference moments in the chosen frame.

Let
\begin{align}\label{eq:manifold}
 \mathcal W&\coloneqq\{w:\empavg{w}=0,\ \empavg{ww^\top}=I_d\},\nonumber\\
 \mathcal Z&\coloneqq\{w\in\mathcal W:E_3=\cdots=E_K=0\}.
\end{align}
The whitening constraints imply $\sum_i\|w_i\|^2=Nd$, so the closed set $\mathcal W$ is bounded and hence compact. We use the empirical norm $\|\cdot\|_N$ to estimate trajectory lengths in the convergence proof.

\begin{thm}[Local convergence]\label{thm:local}
Fix $\bar w\in\mathcal Z$, such as the whitened reference $w^*$.
Suppose each $Q_{m-2}(\bar w)$ has full column rank and each $\mathcal H_m(\bar w)$ is positive definite, $3\le m\le K$ (cf.\ Proposition~\ref{prop:agents}).
For any positive gains $k_m$, there exist a neighborhood $U\subset\mathcal W$ of $\bar w$ and constants $c,\alpha>0$ such that trajectories from $U$ exist for all $t\ge0$, remain in a regular neighborhood, and satisfy
\begin{equation}\label{eq:localestimate}
 \Big(\sum_{m=3}^K\|E_m(t)\|_F^2\Big)^{1/2}
 \le c\,\mathrm e^{-\alpha t}\Big(\sum_{m=3}^K\|E_m(0)\|_F^2\Big)^{1/2}.
\end{equation}
Each trajectory converges to a point $w_\infty\in\mathcal Z$, with $\|w(t)-w_\infty\|_N$ obeying the same type of exponential bound.
\end{thm}
\begin{pf}
Choose $R>0$ so that the compact relative neighborhood
$B\coloneqq\{w\in\mathcal W:\|w-\bar w\|_N\le R\}$ lies in the domain where all $Q_{m-2}$ have full column rank and all $\mathcal H_m$ are positive definite.
Continuity gives uniform bounds $\mathcal H_m\succeq g_m I$ and $\|\mathcal B_{mr}\|\le b_{mr}$ on $B$, with $g_m>0$ and finite $b_{mr}$. Operator norms here and below are induced by the stated domain and range norms.
Write $\zeta_m\coloneqq\|E_m\|_F$ and $a_m\coloneqq mk_mg_m>0$.
For positive weights $\rho_m$ to be chosen, define
\begin{equation*}
 \mathcal V\coloneqq\tfrac12\sum_{m=3}^K\rho_m \zeta_m^2.
\end{equation*}
Differentiating along \eqref{eq:cascade} and applying the operator bounds yields
\begin{equation*}
 \begin{aligned}
 \dot{\mathcal V}
 &=-\sum_{m=3}^K\rho_m mk_m
   \langle E_m,\mathcal H_mE_m\rangle_F\\
 &\quad+\sum_{m=4}^K\sum_{r=3}^{m-1}
   \rho_m\langle E_m,\mathcal B_{mr}E_r\rangle_F\\
 &\le-\sum_{m=3}^K\rho_m a_m \zeta_m^2
   +\sum_{m=4}^K\sum_{r=3}^{m-1}\rho_m b_{mr}\zeta_m\zeta_r.
 \end{aligned}
\end{equation*}
For $m\ge4$, Young's inequality gives, for each $r<m$,
\begin{equation*}
 b_{mr}\zeta_m\zeta_r
 \le\frac{a_m}{2(m-3)}\zeta_m^2+c_{mr}\zeta_r^2,
 \quad c_{mr}\coloneqq\frac{(m-3)b_{mr}^2}{2a_m}.
\end{equation*}
There are $m-3$ such cross terms at level $m$, so together they use at most half of its negative diagonal term.
Collecting the remaining terms by error order gives
\begin{equation*}
 \dot{\mathcal V}\le
 -\sum_{r=3}^K\left(\frac{\rho_ra_r}{2}
 -\sum_{m=r+1}^K\rho_mc_{mr}\right)\zeta_r^2.
\end{equation*}
The triangular structure permits a backward choice of weights: set $\rho_K\coloneqq1$ and, successively for $r=K-1,\ldots,3$, choose
\begin{equation*}
 \rho_r\coloneqq1+\frac{4}{a_r}\sum_{m=r+1}^K\rho_mc_{mr}.
\end{equation*}
Then $\sum_{m>r}\rho_mc_{mr}\le\rho_ra_r/4$ for every $r$ (empty sums are zero).
Consequently, with $\alpha\coloneqq\tfrac14\min_m a_m>0$,
\begin{equation}\label{eq:weighted}
 \dot{\mathcal V}\le-\tfrac14\sum_{m=3}^K\rho_ma_m\zeta_m^2
 \le-2\alpha\mathcal V\quad\text{on }B.
\end{equation}
Thus, as long as the trajectory remains in $B$, $\mathcal V(t)\le \mathrm e^{-2\alpha t}\mathcal V(0)$.
For $\rho_{\min}\coloneqq\min_m\rho_m$ and $\rho_{\max}\coloneqq\max_m\rho_m$,
\begin{equation*}
 \frac{\rho_{\min}}2\sum_m \zeta_m^2\le\mathcal V
 \le\frac{\rho_{\max}}2\sum_m \zeta_m^2,
\end{equation*}
which gives \eqref{eq:localestimate} with $c\coloneqq\sqrt{\rho_{\max}/\rho_{\min}}$.

It remains to show that the trajectory stays in this regular neighborhood. The operators $\mathcal A_m$ are bounded on $B$, so \eqref{eq:hierarchy} and Cauchy--Schwarz give
\begin{equation*}
 \|\dot w\|_N\le\sum_{m=3}^K k_m\|\mathcal A_m\|\zeta_m
 \le L\sqrt{\mathcal V}
\end{equation*}
for some constant $L$.
Up to a first exit from $B$, integration therefore yields
\begin{equation*}
 \|w(t)-w(0)\|_N
 \le\int_0^t\|\dot w(s)\|_N\,ds
 \le\frac{L}{\alpha}\sqrt{\mathcal V(0)}.
\end{equation*}
Since $\mathcal V(\bar w)=0$, choose an open neighborhood $U\subset\mathcal W$ of $\bar w$ such that $\|w(0)-\bar w\|_N<R/2$ and $L\sqrt{\mathcal V(0)}/\alpha<R/4$ for every $w(0)\in U$.
The trajectory then stays within distance $3R/4$ of $\bar w$ and cannot reach the boundary of $B$.
Smoothness of the vector field on a neighborhood of this compact subset gives existence for all $t\ge0$, and the exponential estimates hold globally in time for these initial conditions.
Moreover, $\int_0^\infty\|\dot w(s)\|_N\,ds<\infty$, so $w(t)$ converges to a limit $w_\infty$.
Continuity and $E_m(t)\to0$ imply $w_\infty\in\mathcal Z$, and integrating from $t$ to infinity gives
\begin{equation*}
 \|w(t)-w_\infty\|_N
 \le\frac{L}{\alpha}\sqrt{\mathcal V(0)}\mathrm e^{-\alpha t}.
\end{equation*}

\end{pf}

\begin{rem}[Physical-coordinate convergence]\label{rem:physical-convergence}
Theorem~\ref{thm:local} establishes convergence in whitened coordinates; to recover the physical formation objective, we combine this result with covariance convergence through the reconstruction $p_i=p_c+V\Lambda^{1/2}w_i$.
Under \eqref{eq:physical}, the dispersion controller \eqref{eq:base} ensures
\begin{equation*}
 (\lambda(t))_{a_1}\ge\min\{(\lambda(0))_{a_1},(\lambda^*)_{a_1}\}>0
\end{equation*}
for positive initial and desired covariance eigenvalues, and $\Lambda(t)\to\Lambda^*$ exponentially.
Hence whitening remains well defined, and convergence to $\mathcal Z$ in Theorem~\ref{thm:local}, together with the covariance convergence, implies that the physical configuration approaches the reference moment-similarity class $\mathcal F_K$.
If a desired centroid is specified, \eqref{eq:centroid-control} also gives $p_c(t)\to p_c^*$.
\end{rem}

\begin{rem}[Point-agent collision avoidance]
For the exact continuous-time dynamics in a regular neighborhood, initially distinct point agents remain at distinct positions at every finite time.
All agents follow the same smooth, time-dependent polynomial velocity field, so uniqueness of its trajectories prevents two agents from reaching the same position at the same time.
The invertible transformation back to physical coordinates preserves this property.
This guarantees collision avoidance for agents modeled as points, but does not provide a uniform positive separation distance; sampled or inexact implementations require separate analysis.
\end{rem}

\subsection{Joint higher-moment control and convergence}\label{sec:joint}
The hierarchy uses one gain per order, and its decay rates depend on the spectra of the operators $\mathcal H_m$.
We now regulate the selected higher-order moments jointly through a regularized Gauss--Newton (Levenberg--Marquardt) step with a single parameter $\mu$: an error direction in which the feature Gram matrix has eigenvalue $s$ decays at rate $s/(s+\mu)$, close to one when $s\gg\mu$ (Theorem~\ref{thm:joint-local}). At high orders this matrix becomes ill conditioned, so some directions converge slowly (Section~\ref{sec:numerics}).
The centroid and covariance controllers are retained, and projection preserves their dynamics; the controlled higher-order moments, however, may interact during the transient.
To construct the joint controller, we choose polynomial features whose empirical averages define a joint moment-error vector and whose gradients provide the candidate velocity directions.
We use normalized products of the probabilists' Hermite polynomials $\operatorname{He}_j$, where $j$ denotes the degree, as these features.
Each $\operatorname{He}_j(x)$ has leading term $x^j$ and lower-degree corrections, allowing the features to represent the standardized moment errors $E_3,\ldots,E_K$ in alternative coordinates.
For a multi-index $\alpha\in\mathbb N_0^d$ (a vector of nonnegative integers), write $|\alpha|\coloneqq\sum_{a_1=1}^d(\alpha)_{a_1}$ and $\alpha!\coloneqq\prod_{a_1=1}^d(\alpha)_{a_1}!$.
Here $(\alpha)_{a_1}$ is a polynomial degree in coordinate $a_1$, rather than an agent label.
Define
\begin{equation}\label{eq:hermite}
 \phi_\alpha(w)\coloneqq\frac{\prod_{a_1=1}^d\operatorname{He}_{(\alpha)_{a_1}}((w)_{a_1})}{\sqrt{\alpha!}},
 \quad 3\le|\alpha|\le K,
\end{equation}
where $\operatorname{He}_0(x)\coloneqq1$, $\operatorname{He}_1(x)\coloneqq x$, and $\operatorname{He}_{j+1}(x)\coloneqq x\operatorname{He}_j(x)-j\operatorname{He}_{j-1}(x)$.
Let $\phi(w)$ stack these features in a fixed order, and define $h\coloneqq\empavg{\phi(w)}$, $h^*\coloneqq\empavg{\phi(w^*)}$, and $e\coloneqq h-h^*$. All feature indices and sums below range over the selected multi-indices.
Because the standardized moments of orders zero through two are fixed, $e$ is an invertible triangular transformation of the independent components of $E_3,\ldots,E_K$; hence $e=0$ is equivalent to matching all these moments.
As in the hierarchical construction, we project the feature gradients onto $\mathcal P_1^\perp$ to obtain velocity directions that preserve the centroid and covariance dynamics.
The projected directions and their Gram matrix are
\begin{equation}\label{eq:joint-Gram}
 B_\alpha\coloneqq R_1[\nabla\phi_\alpha(w_i)]_{i=1}^N,\quad
 G_{\alpha\beta}\coloneqq\empavg{B_\alpha\cdot B_\beta}.
\end{equation}
Here $\nabla\phi_\alpha$ is the gradient with respect to $w$, and $B_{i\alpha}\in\mathbb R^d$ is the direction at agent $i$. The dot denotes the Euclidean inner product.
The matrix $G$ is symmetric by definition, and for any coefficient vector $c$ indexed by the selected features,
\begin{equation}\label{eq:joint-regularity}
 \begin{aligned}
 c^\top Gc
 =\sum_{\alpha,\beta}c_\alpha c_\beta
       \empavg{B_\alpha\cdot B_\beta}
 =\frac1N\sum_{i=1}^N
       \left\|\sum_\alpha c_\alpha B_{i\alpha}\right\|_2^2\ge0.
 \end{aligned}
\end{equation}
Thus $G\succeq0$: its quadratic form is the mean squared norm of the combined velocity direction.
For any $\mu>0$, the regularized joint feedback is
\begin{align}\label{eq:joint}
 v_i&\coloneqq-\sum_\alpha B_{i\alpha}[(G+\mu I)^{-1}e]_\alpha,\nonumber\\
 u_i&\coloneqq u_i^{(1)}+u_i^{(2)}+V\Lambda^{1/2}v_i.
\end{align}
The term $\mu I$ ensures invertibility even when $G$ is singular, but does not eliminate its null directions.

\begin{prop}[Joint-moment dynamics and dissipation]\label{prop:joint}
The feedback \eqref{eq:joint} preserves the centroid and covariance equations of $u^{(1)}+u^{(2)}$, and satisfies
\begin{align}\label{eq:joint-decay}
 \dot w_i&=v_i,\quad \dot e=-G(G+\mu I)^{-1}e,\nonumber\\
 \frac{d}{dt}\frac{\|e\|_2^2}{2}
 &=-e^\top G(G+\mu I)^{-1}e\le0.
\end{align}
\end{prop}
\begin{pf}
The components of $v$ are orthogonal to $\mathcal P_1$, so Lemma~\ref{lem:freeze} and the affine cancellation preserve the centroid and covariance equations and give $\dot w=v$.
Differentiating $h_\alpha$ pairs $\nabla\phi_\alpha$ with $v$; its affine part is orthogonal to $v$, giving the stated equation for $e$.
As noted above, the Gram matrix satisfies $G\succeq0$; hence $G(G+\mu I)^{-1}$ is positive semidefinite, proving the dissipation inequality.
\end{pf}

By \eqref{eq:joint-regularity}, $G\succ0$ exactly when the projected feature gradients are linearly independent, and then every nonzero error dissipates strictly.
As for the hierarchy, full column rank of $Q_{K-1}(w^*)$ suffices for $G(w^*)\succ0$ (cf.\ Proposition~\ref{prop:agents}(ii)): if $c^\top G(w^*)c=0$, then by \eqref{eq:joint-regularity} the gradient of $P\coloneqq\sum_\alpha c_\alpha\phi_\alpha$, of degree at most $K-1$, is affine at every agent and hence everywhere, so $P$ is at most quadratic and $c=0$.

\begin{thm}[Local convergence of joint control]\label{thm:joint-local}
Fix $\bar w\in\mathcal Z$ with $G(\bar w)\succ0$ and any $\mu>0$.
There exist a neighborhood $U\subset\mathcal W$ of $\bar w$ and constants $g,\sigma,C>0$, with $\sigma\coloneqq g/(g+\mu)$, such that trajectories of \eqref{eq:joint} from $U$ exist for all $t\ge0$, remain in a neighborhood where $G(w)\succeq gI$, and satisfy
\begin{equation}\label{eq:joint-local-decay}
 \|e(t)\|_2\le \mathrm e^{-\sigma t}\|e(0)\|_2.
\end{equation}
Each trajectory converges to a point $w_\infty\in\mathcal Z$, with
\begin{equation*}
 \|w(t)-w_\infty\|_N\le C\,\mathrm e^{-\sigma t}\|e(0)\|_2.
\end{equation*}
\end{thm}
\begin{pf}
By continuity, choose $r>0$ and $g>0$ such that $G(w)\succeq gI$ on the compact relative neighborhood $B\coloneqq\{w\in\mathcal W:\|w-\bar w\|_N\le r\}$.
On $B$, the eigenvalues of $G(G+\mu I)^{-1}$ are at least $\sigma$, so Proposition~\ref{prop:joint} gives $\frac{d}{dt}\frac12\|e\|_2^2\le-\sigma\|e\|_2^2$, hence \eqref{eq:joint-local-decay} while the trajectory remains in $B$.
With $\eta\coloneqq(G+\mu I)^{-1}e$, the control law \eqref{eq:joint} and \eqref{eq:joint-regularity} give
\begin{equation*}
 \|v\|_N^2=\eta^\top G\eta
 =e^\top(G+\mu I)^{-1}G(G+\mu I)^{-1}e
 \le\frac{\|e\|_2^2}{4\mu},
\end{equation*}
because $s/(s+\mu)^2\le1/(4\mu)$ for $s\ge0$.
The field is smooth on $\mathcal W$, since $G+\mu I$ is invertible.
With this velocity bound in place of $\|\dot w\|_N\le L\sqrt{\mathcal V}$, the trajectory-length argument in the proof of Theorem~\ref{thm:local} keeps trajectories from a suitable neighborhood $U$ inside $B$, gives existence for all $t\ge0$, and yields convergence to some $w_\infty$ with the stated tail, $C\coloneqq(2\sqrt\mu\,\sigma)^{-1}$.
Finally, $e(w_\infty)=0$ is equivalent to $w_\infty\in\mathcal Z$.
\end{pf}

As in Remark~\ref{rem:physical-convergence}, the unchanged centroid and covariance equations transfer Theorem~\ref{thm:joint-local} to the physical moment-similarity objective.

\section{Distributed Moment Control}\label{sec:distributed}
Following \cite{chen2026}, agents reconstruct the shared statistics of Section~\ref{sec:general} from relative measurements and neighbor communication. A centered-position estimator feeds parallel moment estimators, both faster than agent motion. We establish reconstruction, local control evaluation, and slow--fast convergence for both controllers.

\subsection{Distributed estimation of configuration statistics}
Use the fixed connected undirected graph, neighbor sets $\mathcal N_i$, and weights $a_{ij}$ defined in Section~\ref{sec:graph}.
Agent $i$ measures $p_i-p_j$ for $j\in\mathcal N_i$ and exchanges estimator outputs with those neighbors. All agents express relative-position measurements and exchanged quantities in a common coordinate orientation; a common absolute position origin is not required.
Let $R\ge2$ be the highest estimated moment order, chosen in Section~\ref{sec:distributed-implementation}. Let $\theta(z)$ collect the distinct monomials of degrees $2$ through $R$, with dimension $n_R\coloneqq\binom{d+R}{R}-d-1$.
Its empirical mean contains the independent entries of the central moments needed by the controllers.
Each agent maintains $\hat z_i\in\mathbb R^d$ and an auxiliary state $\hat\theta_i\in\mathbb R^{n_R}$, initialized so that
\begin{equation}\label{eq:estimator-init}
 \sum_i\hat z_i(0)=0,\quad \sum_i\hat\theta_i(0)=0.
\end{equation}
Setting all states to zero is one such initialization for reconstruction with fixed positions.
Let $\varepsilon_z>0$ and $\varepsilon_M>0$ denote the time scales of the centered-position and moment estimators, respectively, with fixed ratio $\varepsilon_z=\delta\varepsilon_M$, $0<\delta\le1$.
The two estimator layers are
\begin{align}\label{eq:estimate}
 \varepsilon_z\dot{\hat z}_i
 &=-\sum_{j\in\mathcal N_i}a_{ij}[(\hat z_i-\hat z_j)-(p_i-p_j)],\\
 \varepsilon_M\dot{\hat\theta}_i
 &=-\sum_{j\in\mathcal N_i}a_{ij}[(\hat\theta_i-\hat\theta_j)
 -(\theta(\hat z_i)-\theta(\hat z_j))],\nonumber
\end{align}
The first layer uses the centroid-relative position estimator of \cite{bautista2026source}, with time scale $\varepsilon_z$, to estimate $z_i=p_i-p_c$ by matching relative positions while maintaining zero mean.
The second layer, a dynamic average consensus estimator \cite{kia2019}, updates all required moment components in parallel; it does not require a separate time scale for each moment order.
Taking $\varepsilon_M$ small makes both layers fast relative to the controller, while $\delta<1$ makes the position layer faster than the moment layer.
The auxiliary state $\hat\theta_i$ estimates the local deviation $\theta(z_i)-\empavg{\theta(z)}$.
Its moment estimate is the output
\begin{equation}\label{eq:moment-estimator-output}
 \hat\nu_i\coloneqq \theta(\hat z_i)-\hat\theta_i,
\end{equation}
rather than the auxiliary state $\hat\theta_i$ itself.
Subtracting this deviation yields an estimate of the global moment vector $\empavg{\theta(z)}$.
In particular, the second line of \eqref{eq:estimate} can be evaluated as
$\varepsilon_M\dot{\hat\theta}_i=\sum_{j\in\mathcal N_i}a_{ij}(\hat\nu_i-\hat\nu_j)$.
Thus each neighbor message need only contain $\hat z_i$ and $\hat\nu_i$.

\begin{prop}[Static moment reconstruction]\label{prop:static-estimation}
For fixed agent positions, the connected graph and initialization \eqref{eq:estimator-init} imply
\begin{equation}\label{eq:estimator-limit}
 \lim_{t\to\infty}\hat z_i(t)=p_i-p_c,\quad
 \lim_{t\to\infty}\hat\nu_i(t)=\empavg{\theta(z)}
\end{equation}
for every agent $i$.
\end{prop}
\begin{pf}
Summing either estimator equation over the undirected graph shows that both state sums remain zero.
Using $L_{\mathcal G}$ from Section~\ref{sec:graph}, stack agent vectors as columns, e.g., $z=(z_1^\top,\ldots,z_N^\top)^\top$; use the same convention for estimator states and outputs.
Since the positions are fixed and edge differences of $p$ and $z$ coincide, the first-layer error satisfies
\begin{equation*}
 \varepsilon_z\frac{d}{dt}(\hat z-z)
 =-(L_{\mathcal G}\otimes I_d)(\hat z-z).
\end{equation*}
This error has zero mean, and connectivity makes the Laplacian positive definite on that subspace, so $\hat z(t)\to z$ exponentially as $t\to\infty$.
Set $q_i\coloneqq \theta(z_i)$ and $\bar q\coloneqq N^{-1}\sum_i q_i$.
The zero-mean vector $\hat\theta-(q-\mathbf1\otimes\bar q)$ obeys a stable Laplacian equation driven by
$(L_{\mathcal G}\otimes I_{n_R})([\theta(\hat z_i)]_{i=1}^N-q)$.
This input decays to zero exponentially as $t\to\infty$, because $\theta$ is polynomial and $\hat z(t)$ converges exponentially to $z$.
Consequently, $\lim_{t\to\infty}\hat\theta_i(t)=q_i-\bar q$, and \eqref{eq:moment-estimator-output} gives $\lim_{t\to\infty}\hat\nu_i(t)=\bar q$.
\end{pf}

\subsection{Distributed control implementation}\label{sec:distributed-implementation}
The highest moment order $R$ to be estimated is determined by the quantities needed to evaluate the chosen control law.
For the hierarchy, let $m\in\{3,\ldots,K\}$ denote the moment order controlled at the current level, with $K$ the highest controlled order. Entries of $Q_{m-2}^\top Q_{m-2}$ require moments through order $2m-4$, and the projection right-hand side requires order $2m-3$.
Including $E_m$, level $m$ needs moments through $\max\{m,2m-3\}$, so the complete hierarchy requires $R=\max\{K,2K-3\}$.
For the joint controller, the feature errors and affine projection require moments through order $K$, while the Gram matrix $G$ requires order $2K-2$.
Thus the regularized joint law requires $R=2K-2$; without Gram normalization, a joint gradient law requires only $R=K$.

Agent $i$ arranges the corresponding entries of $\hat\nu_i$ into symmetric central-moment estimates $\hat M_{r,i}$, $2\le r\le R$.
Using the same principal-frame convention and reference tensors as the centralized controller, it computes
\begin{equation}\label{eq:local-whitening}
 \begin{aligned}
 \hat C_i&\coloneqq\hat M_{2,i}=\hat V_i\hat\Lambda_i\hat V_i^\top,\\
 \hat T_i&\coloneqq\hat\Lambda_i^{-1/2}\hat V_i^\top,\quad
 \hat w_i\coloneqq\hat T_i\hat z_i,\\
 \hat S_{r,i}&\coloneqq\hat T_i^{\otimes r}\hat M_{r,i},\quad 2\le r\le R,\\
 \hat E_{r,i}&\coloneqq\hat S_{r,i}-S_r^*,\quad 3\le r\le K.
 \end{aligned}
\end{equation}
The estimated standardized moments supply the projection coefficients and, for the joint law, the Gram entries.
Let $\widehat v_i$ be the value obtained by substituting these estimates into the hierarchical sum $\sum_{m=3}^K v^{(m)}(w_i)$ or the joint law \eqref{eq:joint}.
When no absolute centroid is prescribed, we omit the centroid-control input $u_i^{(1)}$ in \eqref{eq:centroid-control} and implement the distributed control law in physical coordinates as
\begin{equation}\label{eq:distributed-control}
 \begin{aligned}
 \dot p_i=\hat u_i
 &\coloneqq-\hat V_i\operatorname{diag}(\hat\lambda_i-\lambda^*)
       \hat V_i^\top\hat z_i +\hat V_i\hat\Lambda_i^{1/2}\widehat v_i,
 \end{aligned}
\end{equation}
where $\hat\lambda_i$ collects the diagonal entries of $\hat\Lambda_i$.
This avoids reconstructing all rows of $Q_{m-2}$ at every agent: the normal-equation matrices and right-hand sides can be assembled from moment estimates.
These evaluations require positive definite estimated covariance, the relevant projection inverses and, for the joint law, invertibility of the estimated regularized Gram matrix, as well as a consistent choice of eigenvector signs and ordering.
An optional absolute centroid objective additionally requires absolute-position or centroid-reference information; relative positions alone determine only the centered configuration.

The transmitted pair $(\hat z_i,\hat\nu_i)$ has
\begin{equation*}
 d+n_R=\binom{d+R}{R}-1
\end{equation*}
scalar entries.
For $d=2,K=3$, this gives nine for the hierarchy and fourteen for the regularized joint law.

The distributed control law generally perturbs the exact lower-moment preservation identities while the local estimates disagree.
Convergence of the distributed closed-loop system therefore requires analyzing the coupled estimation and motion dynamics in addition to static reconstruction.

\subsection{Local convergence of the slow--fast closed-loop system}
We analyze the slow--fast closed-loop system formed by the distributed estimators \eqref{eq:estimate} and the control law \eqref{eq:distributed-control}.
The main proof uses the hierarchical moment controller, and the joint controller is treated in Corollary~\ref{cor:distributed-joint}.
For fixed $\delta$, the small parameter $\varepsilon_M$ gives singularly perturbed dynamics in which estimation is fast relative to agent motion.
The target is a set of moment-similar configurations, so the proof treats the moment and estimation errors together and bounds the remaining configuration motion.
Smooth local evaluation of the estimated principal frame requires the following assumption, which the centralized controllers do not need.
\begin{assum}\label{ass:distributed}
The graph $\mathcal G$ is fixed, undirected, and connected; the estimators satisfy \eqref{eq:estimator-init}; the reference covariance $C^*$ has distinct positive eigenvalues; and all agents use common eigenvalue-ordering and eigenvector-sign conventions that yield a smooth principal frame near $C^*$.
\end{assum}

Let $\Pi\coloneqq I_N-N^{-1}\mathbf1\mathbf1^\top$ be the centering projector and $q(z)\coloneqq[\theta(z_i)]_{i=1}^N$ the stacked monomial vector. Define the estimation errors
\begin{equation}\label{eq:estimation-errors}
 \tilde z\coloneqq\hat z-z,\quad
 \tilde\theta\coloneqq\hat\theta-(\Pi\otimes I_{n_R})q(z),\quad
 \tilde\xi\coloneqq(\tilde z,\tilde\theta).
\end{equation}
The initialization \eqref{eq:estimator-init} makes both errors zero mean for all time.
Here $\|\tilde\xi\|_N^2\coloneqq N^{-1}\sum_i(\|\tilde z_i\|^2+\|\tilde\theta_i\|^2)$; the individual stacked errors use the corresponding terms.
Writing $L_d\coloneqq L_{\mathcal G}\otimes I_d$ and $L_\theta\coloneqq L_{\mathcal G}\otimes I_{n_R}$, their exact moving-agent dynamics are
\begin{equation}\label{eq:estimation-error-dynamics}
 \begin{aligned}
 \varepsilon_z\dot{\tilde z}&=-L_d\tilde z-\varepsilon_z\dot z,\\
 \varepsilon_M\dot{\tilde\theta}&=-L_\theta\tilde\theta
   +L_\theta\big(q(z+\tilde z)-q(z)\big)\\
 &\quad-\varepsilon_M(\Pi\otimes I_{n_R})Dq(z)\dot z,
 \end{aligned}
\end{equation}
where $Dq(z)$ is the Jacobian of the stacked monomial map and
$\dot z=(\Pi\otimes I_d)\hat u$.
The terms containing $\dot z$ describe the tracking error induced by agent motion.
For frozen positions, these terms vanish and the fast subsystem is a stable cascade on the zero-mean subspaces.
The slow error collects the dispersion and higher-order moment errors,
\begin{equation}\label{eq:slow-error}
 \mathcal E(p)^2\coloneqq\|\lambda(p)-\lambda^*\|^2
       +\sum_{m=3}^K\|E_m(w(p))\|_F^2.
\end{equation}

\begin{thm}[Distributed local convergence]\label{thm:distributed}
Let Assumption~\ref{ass:distributed} hold, let $R=\max\{K,2K-3\}$, and let the whitened reference $w^*$ satisfy the regularity conditions of Theorem~\ref{thm:local}.
For any positive gains, there exist $\varepsilon_0,c,\alpha>0$ such that, for $0<\varepsilon_M\le\varepsilon_0$, the estimators \eqref{eq:estimate} with the hierarchical realization of \eqref{eq:distributed-control}, started near $(p^*,\tilde\xi=0)$, are well defined for all $t\ge0$ and satisfy
\begin{equation}\label{eq:distributed-decay}
 \mathcal E(p(t))+\|\tilde\xi(t)\|_N
 \le c\,\mathrm e^{-\alpha t}\big(\mathcal E(p(0))+\|\tilde\xi(0)\|_N\big).
\end{equation}
Moreover, $p(t)$ converges exponentially to some $p_\infty\in\mathcal F_K$, and every moment estimate converges to the true central moment.
\end{thm}
\begin{pf}
Choose a compact neighborhood of $(p^*,0)$ on which the true and estimated covariance matrices have uniformly positive, distinct eigenvalues and the projection inverses exist.
The control and moment maps are smooth with bounded derivatives on this neighborhood.
We derive the following estimates up to the first exit time and subsequently establish that sufficiently close initial conditions preclude such an exit.

At $\tilde\xi=0$, \eqref{eq:distributed-control} equals the centralized physical hierarchical field, denoted by $u_{\mathrm{ex}}(p)$.
Let $\mathcal V(w)$ be the weighted moment energy constructed in Theorem~\ref{thm:local}, and set
\begin{equation*}
 V_s(p)\coloneqq\tfrac12\|\lambda(p)-\lambda^*\|^2+\mathcal V(w(p)).
\end{equation*}
The exact covariance dynamics and \eqref{eq:weighted} give
$\dot V_s|_{u_{\mathrm{ex}}}\le-c_s\mathcal E^2$ for some $c_s>0$.
Moreover, $V_s$ is bounded above and below by positive multiples of $\mathcal E^2$, and its gradient has norm at most a constant times $\mathcal E$.
Smooth dependence of the estimated controller on its inputs gives
\begin{equation*}
 \begin{aligned}
 \|\hat u-u_{\mathrm{ex}}(p)\|_N&\le C\|\tilde\xi\|_N,\\
 \|\hat u\|_N+\|\dot z\|_N&\le C(\mathcal E+\|\tilde\xi\|_N),\\
 \dot V_s&\le-c_s\mathcal E^2+C_s\mathcal E\|\tilde\xi\|_N.
 \end{aligned}
\end{equation*}
Here the velocity bound uses $u_{\mathrm{ex}}=0$ whenever $\mathcal E=0$ and the explicit linear dependence of the feedback on its errors.

For the estimator, write $\ell\coloneqq\lambda_2(L_{\mathcal G})>0$.
The polynomial map is locally Lipschitz, so for some $M$,
$\|L_\theta(q(z+\tilde z)-q(z))\|_N\le M\|\tilde z\|_N$.
Choose $\beta>0$ so that $\beta\ell/\delta>M^2/(2\ell)$ and define
\begin{equation*}
 V_f\coloneqq\tfrac\beta2\|\tilde z\|_N^2+\tfrac12\|\tilde\theta\|_N^2.
\end{equation*}
Using the zero-mean property in \eqref{eq:estimation-error-dynamics}, the relation $\varepsilon_z=\delta\varepsilon_M$, and Young's inequality for the cross term yields
\begin{equation*}
 \begin{aligned}
 \dot V_f
 &\le-\frac{1}{\varepsilon_M}\left[
 \left(\frac{\beta\ell}{\delta}-\frac{M^2}{2\ell}\right)
 \|\tilde z\|_N^2+\frac\ell2\|\tilde\theta\|_N^2\right]\\
 &\quad+C_f\|\tilde\xi\|_N\|\dot z\|_N\\
 &\le-\left(\frac{c_f}{\varepsilon_M}-C_1\right)\|\tilde\xi\|_N^2
       +C_2\mathcal E\|\tilde\xi\|_N
 \end{aligned}
\end{equation*}
for $c_f>0$ and finite nonnegative constants $C_f,C_1,C_2$.
In particular, frozen positions give uniform exponential stability of the fast error system.

For the composite energy $V_{\mathrm{tot}}\coloneqq V_s+V_f$, another application of Young's inequality gives
\begin{equation*}
 \dot{V}_{\mathrm{tot}}\le-\tfrac{c_s}{2}\mathcal E^2
 -\left(\frac{c_f}{\varepsilon_M}-C_1
          -\frac{(C_s+C_2)^2}{2c_s}\right)\|\tilde\xi\|_N^2.
\end{equation*}
Choose $\varepsilon_0>0$ small enough that the coefficient in parentheses is at least one for all $0<\varepsilon_M\le\varepsilon_0$.
Since $V_{\mathrm{tot}}$ is equivalent to $\mathcal E^2+\|\tilde\xi\|_N^2$, this proves $\dot V_{\mathrm{tot}}\le-2\alpha V_{\mathrm{tot}}$ for a uniform $\alpha>0$ and hence \eqref{eq:distributed-decay} while the trajectory remains in the neighborhood.

Finally, $\|\dot p\|_N\le C\sqrt{V_{\mathrm{tot}}}$ implies
\begin{equation*}
 \int_0^t\|\dot p(s)\|_N\,ds
 \le\frac{C}{\alpha}\sqrt{V_{\mathrm{tot}}(0)}.
\end{equation*}
Choose the initial configuration in a smaller interior neighborhood and $V_{\mathrm{tot}}(0)$ sufficiently small.
The last bound prevents the configuration from reaching its boundary, while $V_f\le V_{\mathrm{tot}}(0)$ keeps the estimator errors inside their chosen neighborhood.
Thus a first exit is impossible, the regularity conditions persist, and smoothness gives global-in-time existence for these local initial conditions.
The velocity is integrable, so $p(t)\to p_\infty$; its exponentially decaying tail gives exponential convergence.
As $\mathcal E\to0$, the limit lies in $\mathcal F_K$.
Finally, \eqref{eq:moment-estimator-output} and \eqref{eq:estimation-errors} give
$\hat\nu_i-\empavg{\theta(z)}
=\theta(z_i+\tilde z_i)-\theta(z_i)-\tilde\theta_i$,
which tends to zero exponentially.
\end{pf}

Thus, when estimation is sufficiently fast relative to motion, the distributed implementation inherits the local exponential convergence of the centralized hierarchy in Theorem~\ref{thm:local}.

\begin{rem}[Time-scale separation in practice]\label{rem:timescale}
The estimation lag scales with $\varepsilon_M/\lambda_2$ times the speed of the agents, and its effect on the control grows with the moment orders that the controller uses.
High orders on sparse graphs therefore demand the fastest communication, whereas regularization, such as $\mu$ in the joint law, reduces the sensitivity to estimation errors (Section~\ref{sec:numerics}).
Since the lag is driven by motion, communication can be reduced as the swarm slows down, for instance with event-triggered dynamic average consensus \cite{kia2019,nowzari2019}.
Under the hierarchy, moreover, the converged orders below the active level stay constant, so their estimates can be frozen, although the projections still need the uncontrolled orders up to $2m-3$.
We focus here on the core mechanism and leave such communication-saving schemes to implementations on robots.
\end{rem}

\begin{cor}[Distributed joint moment control]\label{cor:distributed-joint}
Let Assumption~\ref{ass:distributed} hold, $G(w^*)\succ0$, $\mu>0$, and $R=2K-2$.
Then the conclusions of Theorem~\ref{thm:distributed} hold for the joint realization of \eqref{eq:distributed-control}, with possibly different constants and neighborhood.
\end{cor}
\begin{pf}
Use the slow energy
\begin{equation*}
 V_s(p)\coloneqq\tfrac12\|\lambda(p)-\lambda^*\|^2
             +\tfrac12\|e(w(p))\|_2^2.
\end{equation*}
Since $e$ is an invertible transformation of the independent moment errors, $V_s$ is equivalent to $\mathcal E^2$.
Locally, $G\succeq gI$ for some $g>0$, so Proposition~\ref{prop:joint} gives $\dot V_s|_{u_{\mathrm{ex}}}\le-c_s\mathcal E^2$.
The velocity bound in Theorem~\ref{thm:joint-local} yields $\|u_{\mathrm{ex}}\|_N\le C\mathcal E$.
Continuity preserves estimated regularized-Gram invertibility, giving smooth control evaluation and $\|\hat u-u_{\mathrm{ex}}\|_N\le C\|\tilde\xi\|_N$.
Consequently, $\dot V_s\le-c_s\mathcal E^2+C_s\mathcal E\|\tilde\xi\|_N$.
The unchanged estimator-error structure therefore permits the same composite-energy and trajectory-length argument as Theorem~\ref{thm:distributed}, with adjusted constants.
\end{pf}

Both results require zero estimator-state sums and small estimation errors; zero initialization alone need not give admissible covariance estimates.
Near a regular reference, agents may first hold position while the estimators converge, then activate control once the estimates enter the required neighborhood.
The common-field collision-avoidance argument does not extend to differing local estimates.

With absolute positions and a common desired centroid $p_c^*$, add
\begin{equation}\label{eq:estimated-centroid-control}
 \hat u_i^{(1)}\coloneqq-k_c(p_i-\hat z_i-p_c^*)
\end{equation}
to \eqref{eq:distributed-control}.
At exact estimation, $\dot p_c=-k_c(p_c-p_c^*)$.
Adding $\tfrac12\|p_c-p_c^*\|^2$ to $V_s$ and the centroid error to the total error norm extends the proof: the added dissipation is $-k_c\|p_c-p_c^*\|^2$, and the perturbation $k_c\tilde z_i$ is bounded by the estimation error.

\section{Numerical Studies}\label{sec:numerics}
\begin{figure*}[!t]
 \centering
 \includegraphics[width=\textwidth]{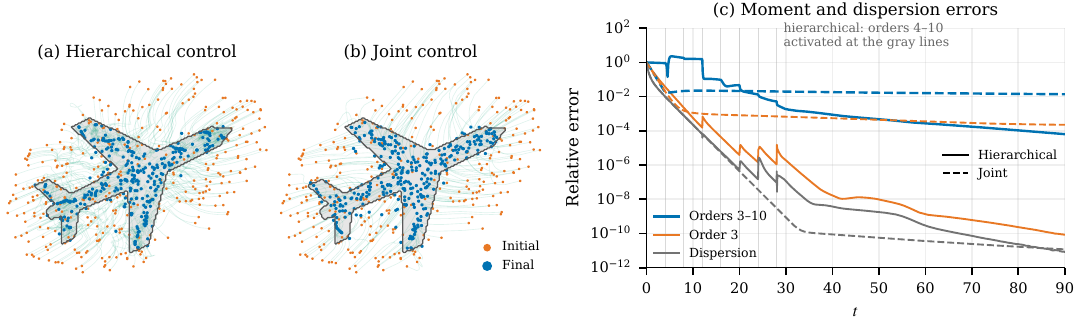}
 \caption{Distributed airplane formation with $N=350$ agents, moments through order $K=10$, and a fixed $6$-nearest-neighbor graph. (a) Hierarchical and (b) joint control: initial agents (orange), trajectories (green), final agents (blue), and target outline. (c) Errors normalized by their initial values: orders $3$ to $10$ combined, order $3$, and dispersion; solid lines are hierarchical, dashed lines joint. The joint law decreases its feature error $e$ (Proposition~\ref{prop:joint}), a different norm of the same moment errors, so its combined error need not decrease monotonically. Gray lines mark the activation of orders $4$ to $10$ in the hierarchy; the brief rises at these instants stem from the discretized, distributed implementation of the switch.}
 \label{fig:airplane-comparison}
\end{figure*}
We compare the hierarchical and joint controllers on the airplane of Fig.~\ref{fig:concept}, in one distributed scenario in which only the higher-moment controller differs.
The $N=350$ agents start from an unstructured filled ellipse, with $19\%$ of them inside the target outline, and regulate moments through order $K=10$ using relative positions only ($u^{(1)}=0$).
They communicate over a fixed $6$-nearest-neighbor graph ($\lambda_2=0.023$), and the estimators use $\varepsilon_M=10^{-9}\lambda_2$.
The hierarchical gains $k_m$ set the geometric mean of the extreme eigenvalues of $mk_m\mathcal H_m(w^*)$ to one.
Since the high-order errors of an unstructured swarm are initially much larger than the low-order ones, the hierarchy builds the formation from coarse to fine, activating order $m$ at $t=4(m-3)$; by Theorem~\ref{thm:hierarchy}, each new order leaves the previous ones unchanged.
The joint law uses $\mu=0.03$ on all orders from the start.

Both swarms form the airplane without agent-to-reference assignments (Fig.~\ref{fig:airplane-comparison}). The joint law places $85\%$ of the agents inside the outline by $t=8$ and ends at $87\%$; the hierarchy, which adds orders until $t=28$, reaches $85\%$ at $t=20$ and $90\%$ at $t=57$, and ends at $91\%$.
The hierarchy reaches the full specification: the combined error of orders $3$ to $10$ falls to $6.4\times10^{-5}$ of its initial value, while the dispersion and order-$3$ errors keep decreasing as orders are added.
The joint law reduces the combined error to about $2\%$ by $t=10$ and then decreases slowly, to $1.4\%$ at $t=90$.
The distributed control stays close to the exact centralized field: within a relative $2\times10^{-8}$ for the joint law and $10^{-5}$ for the hierarchy, whose deviation grows as higher orders are activated (Remark~\ref{rem:timescale}).
\begin{table}[!t]
\caption{Hierarchical versus joint control in Fig.~\ref{fig:airplane-comparison}}
\label{tab:comparison}
\centering
\footnotesize
\setlength{\tabcolsep}{3pt}
\begin{tabular}{@{}lcc@{}}
\toprule
 & Hierarchical & Joint \\
\midrule
Parameters & $8$ gains, schedule & $\mu$ \\
Orders preserved & all lower orders & $1$ and $2$ \\
Conditioning at $w^*$ & $\operatorname{cond}\mathcal H_m\le404$ & $\operatorname{cond}G=5\times10^{8}$ \\
Scalars per message & $170$ & $189$ \\
Final error, orders $3$--$10$ & $6.4\times10^{-5}$ & $1.4\times10^{-2}$ \\
Deviation from exact control & $10^{-5}$ & $2\times10^{-8}$ \\
Agents inside outline & $91\%$ & $87\%$ \\
Peak speed / mean path & $11.2$ / $1.58$ & $4.2$ / $0.86$ \\
\bottomrule
\end{tabular}
\end{table}

Table~\ref{tab:comparison} summarizes the trade-off; both controllers cost about the same to implement.
The hierarchy treats one order at a time, and each of these subproblems is well conditioned, so it reaches the full specification; the price is more motion, with speed peaks when orders are activated, and a higher sensitivity to estimation errors, which calls for faster communication.
The joint law needs a single parameter and moves the swarm gently, but treating all orders at once makes its problem poorly conditioned, so part of the error decays slowly.
Retuning $\mu$ does not close this gap: a smaller value speeds up the slow part only by raising the velocities.
Since a moment error of about $1\%$ already reproduces the silhouette, the joint law suits close approximations, whereas the hierarchy is preferable when the moments themselves are the specification or when the formation is refined by adding orders.

\section{Conclusion}\label{sec:conclusion}
This paper specified formations through spatial moments: the centroid and covariance set the position and principal spreads, and higher-order moments in whitened coordinates encode the shape of a target image, without assigning agents to destinations.

Orthogonalizing in the swarm's own empirical inner product makes each hierarchical correction preserve all lower-order moments exactly. The error dynamics are therefore lower triangular and converge locally exponentially for any positive gains, and a formation can be refined by adding orders.
A joint alternative regulates the higher moments together through a regularized Gauss--Newton step while preserving the centroid and covariance dynamics.
Regularity requires enough well-spread agents; in the plane, full third-order regularity needs at least five, a bound attained by a regular pentagon.
A two-layer estimator based on relative positions and neighbor communication retains local convergence when estimation is fast relative to motion.

In a distributed simulation on a sparse graph, $350$ agents formed an airplane through moments up to order ten. Built from coarse to fine, the hierarchy reached the full specification, whereas the joint law approached it with smaller velocities and converged slowly in the poorly conditioned directions of its Gram matrix.

This paper focused on the core mechanism. Implementations on robots will benefit from communication-saving schemes, such as event-triggered estimation, and from inter-agent separation. Further theoretical steps are regions of attraction beyond the local analysis, systematic activation schedules and gain selection at high orders, and the choice of compact moment sets for a given spatial pattern, since finite-order moments specify a class of configurations rather than a unique density.

\bibliographystyle{IEEEtran}
\bibliography{references}

\end{document}